\documentclass[english]{achemso}
\usepackage[T1]{fontenc}
\usepackage[latin9]{inputenc}
\usepackage{geometry}
\usepackage{amsmath}
\usepackage{amssymb}
\usepackage{graphicx}
\usepackage[numbers]{natbib}

\makeatletter

\title{Enantioselective Topological Frequency Conversion}

\author{Kai Schwennicke and Joel Yuen-Zhou}

\affiliation{Department of Chemistry and Biochemistry, University of California,
San Diego, La Jolla, California 92093, United States}

\email{joelyuen@ucsd.edu}

\newcommand*{\citen}[1]{%
  \begingroup
    \romannumeral-`\x 
    \setcitestyle{numbers}%
    \cite{#1}%
  \endgroup   
}

\@ifundefined{showcaptionsetup}{}{%
 \PassOptionsToPackage{caption=false}{subfig}}
\usepackage{subfig}
\makeatother

\usepackage{babel}
\begin{document}
\captionsetup[figure]{justification=raggedright}

\captionsetup[subfigure]{position=top,justification=raggedright,singlelinecheck=off}
\begin{abstract}
Two molecules are enantiomers if they are non-superimposable mirror
images of each other. Electric dipole-allowed cyclic transitions $|1\rangle\to|2\rangle\to|3\rangle\to|1\rangle$
obey the symmetry relation $\mathcal{O}^{R}=-\mathcal{O}^{S}$, where
$\mathcal{O}^{R,S}=(\mu_{21}^{R,S}E_{21})(\mu_{13}^{R,S}E_{13})(\mu_{32}^{R,S}E_{32})$,
and $R,S$ label the two enantiomers. Here we generalize the concept
of topological frequency conversion to an ensemble of enantiomers.
We show that, within a rotating-frame, the pumping power between fields
of frequency $\omega_{1}$ and $\omega_{2}$ is sensitive to enantiomeric
excess, $\mathcal{P}_{2\to1}=\hbar\frac{\omega_{1}\omega_{2}C_{L}^{R}}{2\pi}(N_{R}-N_{S})$,
where $N_{i}$ is the number of enantiomers $i$ and $C_{L}^{R}$
is an enantiomer-dependent Chern number. Connections with chiroptical
microwave spectroscopy are made. Our work provides an underexplored
and fertile connection between topological physics and molecular chirality.

\end{abstract}
In the mid-nineteenth century, Louis Pasteur discovered that molecules
can possess handedness, or chirality, an attribute that influences
how they interact with their surroundings \citep{pasteur1848relations}.
More generally, the two species of a chiral molecule, referred to
as enantiomers, are nonsuperimposable mirror images of each other
and, while they feature many identical physicochemical properties
(up to very small parity violation corrections \citep{quack2008high}),
they can also exhibit drastically different behavior when exposed
to chiral environments or stimuli. Thus, enantioselectivity plays
a crucial role in biological activity as well as in the synthesis,
purification, and characterization of pharmaceuticals \citep{hutt1996drug,kasprzyk2010pharmacologically,blackmond2010origin}.
Traditionally, optical rotation and circular dichroism have served
as optical tools to obtain enantioselective information; however,
these techniques rely on the weak interaction between molecules and
the magnetic component of the optical field. A very active effort
in chirality research consists of spatially shaping electromagnetic
fields \citep{tang2010optical,poulikakos2018chiral,garcia2018enantiomer}
to enhance these weak interactions. Other techniques that rely solely
on electric dipole interactions \citep{ordonez2018generalized} have
been recently advocated. For instance, many efforts are currently
invested in photoelectron circular dichroism (PECD) \citep{ritchie1976theory,bowering2001asymmetry,lux2012circular,demekhin2018photoelectron}.
Yet, others focus on nonlinear optical signals that depend on the
sign of the electric fields with which the molecules interact \citep{fischer2000three,brumer2001principles},
including photoexcitation circular dichroism \citep{beaulieu2018photoexcitation},
the use of synthetic chiral fields \citep{neufeld2019ultrasensitive,ayuso2019synthetic,neufeld2020degree,ayuso2021enantio},
and microwave three-wave mixing \citep{patterson2013enantiomer,patterson2013sensitive,shubert2014identifying,lobsiger2015molecular,shubert2016chiral}.
More precisely, the latter technique can be understood through cyclic
three-level models \citep{kral2001cyclic,kral2003two,thanopulos2003theory,li2008dynamic,vitanov2019highly,kang2020effective,wu2020two}
where the product of three light-matter couplings {[}hereafter referred
to as the Král-Shapiro (KS) product{]} differs by a phase of $\pi$
between the two enantiomers. This remarkable symmetry has been exploited
to propose cyclic population transfer schemes \citep{kral2001cyclic,wu2020two}
or the use of cross-polarized terahertz pulses \citep{tutunnikov2021enantioselective}
to prepare the enantiomers in different energy configurations or orientations
for separation. This symmetry has also been utilized to suggest an
enantioselective generalization of the Stern-Gerlach \citep{gerlach1922experimentelle}
or spin Hall \citep{hirsch1999spin} experiments, where spatial separation
of enantiomers, rather than spins, is achieved using artificial gauge
fields \citep{li2007generalized,li2010theory,chen2020enantio}. The
analogy between enantiomer and spin labels is intriguing and surprisingly
underexplored, and serves as the motivation of our present work. More
specifically, we wish to demonstrate an enantioselective analogue
to the Quantum Spin Hall Effect (QSHE) \citep{kane2005quantum}.

On the other hand, since the pioneering work of Thouless, Kohmoto,
Nightingale, and den Nijs in relation to the Quantum Hall Effect (QHE)
\citep{thouless1982quantized}, notions of symmetry-protected topological
phases (SPTPs) have been at the heart of condensed matter research,
and have only been exacerbated in the past fifteen years with the
discovery of topological insulators \citep{hasan2010colloquium}.
These notions guarantee that certain response properties of so-called
topologically nontrivial systems are largely independent of material
specification, instead depending only on products of universal constants
and integer quantities known as topological invariants. The discrete
nature of these properties implies that they are robust against material
imperfections, thus making them attractive for metrology, among other
applications. While topological protection was originally identified
in translationally invariant 2D systems, its scope has been enlarged
through the use of Floquet engineering in systems of different dimensionality
\citep{kolodrubetz2018topological,mondragon2018quantized,nathan2019topological,oka2019floquet}
and the consideration of the 2D phase space of 1D systems \citep{leboeuf1990phase,leboeuf1992topological}.
Of particular interest is an elegant construction due to Martin, Refael,
and Halperin \citep{martin2017topological} called topological frequency
conversion (TFC), where quantized \textquotedbl current\textquotedbl{}
is observed. In this Letter, we design a novel spectroscopic scheme
that generalizes TFC to the microwave spectroscopy of an ensemble
of chiral molecules. The very first link between chiroptical spectroscopy
and topology was suggested recently in work by Ordoñez and Smirnova
\citep{ordonez2019propensity} within the context of PECD. These authors
showed that the propensity field (a pseudoscalar) as a function of
ejected photoelectron direction (Berry curvature) can be integrated
over all solid angles to yield a quantized enantiosensitive flux which
is proportional to a Chern number. Similarly, the authors showed that
a microwave three-wave mixing signals can be interpreted in terms
of an analogous quantity to the propensity field \citep{ordonez2018generalized}.
However, it is not clear from that work if there exists a parameter
space upon which integration of the signals lead to topological invariants,
so geometric and topological consequences of these nonlinear spectroscopies
were not explored. In this Letter, we use TFC to identify time as
the missing parameter space and for simplicity, restrict our attention
to frequency conversion rather than three-wave mixing. The result
is a signal that is proportional to enantiomeric excess (EE), with
a simple prefactor containing the sign of the KS product. Owing to
the topological nature of the signal, it should also serve as a very
sensitive detection of EE. As far as we are aware, our work provides
the first connection between topological physics, chiroptical spectroscopy,
and nonlinear spectroscopy, and anticipates a fertile ground for further
exploration.

\begin{figure}
\subfloat[]{\includegraphics[scale=0.9]{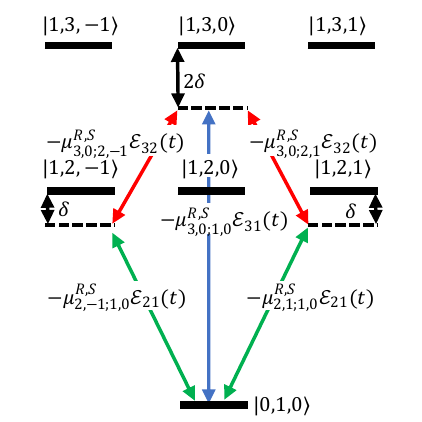}}

\subfloat[]{\includegraphics{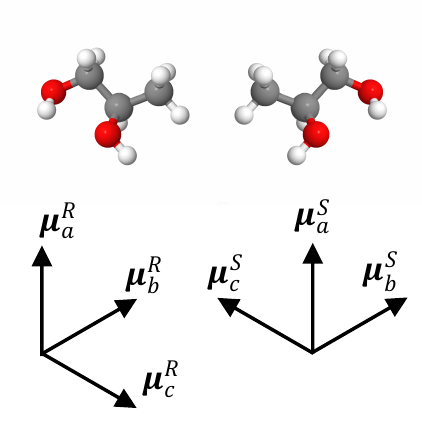}

}\caption{\emph{\small{}The model.}{\small{} (a) Cyclic three-level transitions
for an asymmetric top, such as enantiomers. Three near-resonant, linear
polarized lasers with modulated field amplitudes $\mathcal{E}_{ij}(t)$
interact with these transitions. (b) The principal axes components
of the dipole moments for the $R-$ and $S-$1,2-propanediol enantiomers.
Note that $(\boldsymbol{\mu}_{a}^{R}\cdot\hat{\boldsymbol{a}})(\boldsymbol{\mu}_{b}^{R}\cdot\hat{\boldsymbol{b}})(\boldsymbol{\mu}_{c}^{R}\cdot\hat{\boldsymbol{c}})=-(\boldsymbol{\mu}_{a}^{S}\cdot\hat{\boldsymbol{a}})(\boldsymbol{\mu}_{b}^{S}\cdot\hat{\boldsymbol{b}})(\boldsymbol{\mu}_{c}^{S}\cdot\hat{\boldsymbol{c}})$
}\label{fig:Model-cyclical-three}}
\end{figure}

Following the principles of enantioselective microwave three wave-mixing
\citep{lehmann2018influence,leibscher2019principles}, we treat the
enantiomers as asymmetric tops whose Hamiltonian is 
\begin{equation}
H_{0}=AJ_{a}^{2}+BJ_{b}^{2}+CJ_{c}^{2},\label{eq:asymmetric_top_hamiltonian}
\end{equation}
where $J_{a}$, $J_{b}$, $J_{c}$ are the angular momentum operators
with respect to the principal axes $\hat{\boldsymbol{a}}$, $\hat{\boldsymbol{b}}$,
$\hat{\boldsymbol{c}}$, and $A>B>C$ are the corresponding rotational
constants. The eigenstates are labeled as $|J,\tau,M\rangle$, where
$J=0,1,2...$ is the rotational quantum number, $M=-J,-J+1,-J+2,...,J$
is the quantum number that characterize the projection of the total
angular momentum along the $z$-laboratory-fixed axis, and $\tau$
serves as the quantum number to differentiate between states with
the same $J$ and $M$. We consider the the following low angular
momentum eigenstates of Eq. \ref{eq:asymmetric_top_hamiltonian} with
a rotational quantum number of $J=0$ or $J=1$
\begin{align}
|0,\tau=1,0\rangle & ,\nonumber \\
|1,\tau=2,M\rangle & ,\nonumber \\
|1,\tau=3,M\rangle & ,\label{eq:asymmetric_eigenstates}
\end{align}

\noindent where $M=-1,0,1$ \citep{simons2003introduction} (see Supporting
Information Section 1, SI-1). The ground state $|0,\tau=1,0\rangle$,
with energy $\hbar\epsilon_{1}$, and the excited states $|1,\tau=2,M\rangle$
and $|1,\tau=3,M\rangle$, with energies $\hbar\epsilon_{2}$, $\hbar\epsilon_{3}$,
respectively, are coupled to each other using a set of three orthogonally-polarized
time-dependent electric fields 
\begin{align}
\boldsymbol{\mathbb{E}}_{21}(t) & ={\cal E}_{21}(t)\sin(\Omega_{21}t)\hat{\boldsymbol{y}},\nonumber \\
\boldsymbol{\mathbb{E}}_{32}(t) & =\mathcal{E}_{32}(t)\cos(\Omega_{32}t)\hat{\boldsymbol{x}},\nonumber \\
\boldsymbol{\mathbb{E}}_{31}(t) & =\mathcal{E}_{31}(t)\cos(\Omega_{31}t)\boldsymbol{\hat{\boldsymbol{z}}},\label{eq:electric_fields}
\end{align}
where $\hat{\boldsymbol{x}}$, $\hat{\boldsymbol{y}}$, \textbf{$\hat{\boldsymbol{z}}$}
denote the three laboratory-fixed axes, the frequencies $\Omega_{21}=\epsilon_{2}-\epsilon_{1}-\delta$,
$\Omega_{32}=\epsilon_{3}-\epsilon_{2}-\delta$, $\Omega_{31}=\epsilon_{3}-\epsilon_{1}-2\delta$,
are slightly detuned from the system's natural frequencies, and the
field amplitudes $\mathcal{E}_{21}(t)$, $\mathcal{E}_{32}(t)$, $\mathcal{E}_{31}(t)$
are slowly modulated. Note from the selection rules for electric dipole
interactions \citep{leibscher2019principles}, that $\Delta M=0$
for the $z$ polarized field and $\Delta M=\pm1$ for the $x$ and
$y$ polarized field (see Fig. \ref{fig:Model-cyclical-three}). Ignoring
all states that are not coupled through the driving electric fields,
and assuming that $|\mu_{i,M';j,M}^{R,S}\mathcal{E}_{ij}(t)|/2\ll\hbar\Omega_{ij}$,
the Hamiltonian for the laser dressed $R-$ and $S-$enantiomer, after
making the rotating wave approximation, is 
\begin{align}
H^{R,S}(t) & =\sum_{i=1,3}\hbar\epsilon_{i}|i,0\rangle\langle i,0|+\hbar\epsilon_{2}\sum_{M=\pm1}|2,M\rangle\langle2,M|\nonumber \\
 & -\mathcal{E}_{21}(t)\sum_{M=\pm1}\Big(\frac{i\mu_{2,M;1,0}^{R,S}e^{-i\Omega_{21}t}}{2}|2,M\rangle\langle1,0|+\text{h.c.}\Big)\nonumber \\
 & -\mathcal{E}_{32}(t)\sum_{M=\pm1}\Big(\frac{\mu_{3,0;2,M}^{R,S}e^{-i\Omega_{32}t}}{2}|3,0\rangle\langle2,M|+\text{h.c.}\Big)\nonumber \\
 & -\mathcal{E}_{31}(t)\Big(\frac{\mu_{3,0;1,0}^{R,S}e^{-i\Omega_{31}t}}{2}|3,0\rangle\langle1,0|+\text{h.c.}\Big),\label{eq:Hamiltonian}
\end{align}
where for simplicity we have introduced the notation $|1,0\rangle\equiv|0,\tau=1,0\rangle$,
$|2,M\rangle\equiv|1,\tau=2,M\rangle$, $|3,0\rangle\equiv|1,\tau=3,0\rangle$.
In Eq. \ref{eq:Hamiltonian} $\mu_{i,M';j,M}^{R,S}$ is the component
of the transition-dipole moment for the $|j,M\rangle\to|i,M'\rangle$
transition that is projected along the polarization axis of $\boldsymbol{\mathbb{E}}_{ij}(t)$
. Following the procedure of Refs. \citen{lehmann2018influence,leibscher2019principles},
the values of $\mu_{i,M';j,M}^{R,S}$ are
\begin{align}
\mu_{2,\pm1;1,0}^{R,S} & =-\frac{i\mu_{b}^{R,S}}{\sqrt{6}},\nonumber \\
\mu_{3,0;2,\pm1}^{R,S} & =\frac{\mu_{a}^{R,S}}{2\sqrt{2}},\nonumber \\
\mu_{3,0;1,0}^{R,S} & =-\frac{i\mu_{c}^{R,S}}{\sqrt{3}},\label{eq:dipole_moments}
\end{align}
where $\mu_{a}^{R,S}$,$\mu_{b}^{R,S}$,$\mu_{c}^{R,S}$ are the components
of the dipole moment along the principal molecular axes. These components
are real valued and $|\mu_{a}^{R}|=|\mu_{b}^{S}|$. Note that for
the chosen polarizations for the three electric fields (see Eq. \ref{eq:electric_fields})
and studied energy levels, $\mu_{i,M';j,M}^{R,S}$ does not depend
on the quantum number $M$. The associated time-dependent wavefunction
for the $R-$ and $S-$ enantiomer the system is $|\psi^{R,S}(t)\rangle$.

Next, we consider the rotating frame \begin{small}
\begin{equation}
U(t)=e^{-i(\epsilon_{2}-\Omega_{21})t}|1,0\rangle\langle1,0|+\sum_{M=\pm1}e^{-i\epsilon_{2}t}|2,M\rangle\langle2,M|+e^{-i(\epsilon_{2}+\Omega_{32})t}|3,0\rangle\langle3,0|,\label{eq:rotating_frame}
\end{equation}
\end{small}such that $|\psi^{R,S}(t)\rangle=U(t)|\tilde{\psi}^{R,S}(t)\rangle$,
in order to remove the central frequencies $\Omega_{ij}$. In this
frame, $i\hbar\partial_{t}|\tilde{\psi}{}^{R,S}(t)\rangle=\mathcal{H}^{R,S}(t)|\tilde{\psi}^{R,S}(t)\rangle$,
with the effective Hamiltonian: \begin{small}

\begin{equation}
\mathcal{H}^{R,S}(t)=\frac{1}{2}\begin{pmatrix}-2\hbar\delta & -\frac{\mu_{b}^{R,S}}{\sqrt{6}}\mathcal{E}_{21}(t) & -\frac{\mu_{b}^{R,S}}{\sqrt{6}}\mathcal{E}_{21}(t) & -\frac{i\mu_{c}^{R,S}}{\sqrt{3}}\mathcal{E}_{31}(t)\\
-\frac{\mu_{b}^{R,S}}{\sqrt{6}}\mathcal{E}_{21}(t) & 0 & 0 & -\frac{\mu_{a}^{R,S}}{2\sqrt{2}}\mathcal{E}_{32}(t)\\
-\frac{\mu_{b}^{R,S}}{\sqrt{6}}\mathcal{E}_{21}(t) & 0 & 0 & -\frac{\mu_{a}^{R,S}}{2\sqrt{2}}\mathcal{E}_{32}(t)\\
\frac{i\mu_{c}^{R,S}}{\sqrt{3}}\mathcal{E}_{31}(t) & -\frac{\mu_{a}^{R,S}}{2\sqrt{2}}\mathcal{E}_{32}(t) & -\frac{\mu_{a}^{R,S}}{2\sqrt{2}}\mathcal{E}_{32}(t) & 2\hbar\delta
\end{pmatrix}.\label{eq:rotating_frame_hamiltonian}
\end{equation}
 \end{small} After a change of basis (see SI-2), we arrive at the
following effective Hamiltonian 
\begin{align}
\mathcal{H}^{R,S}(t) & =-\frac{\mu_{b}^{R,S}\mathcal{E}_{21}(t)}{2\sqrt{3}\hbar}L_{x}-\frac{\mu_{a}^{R,S}\mathcal{E}_{32}(t)}{4\hbar}L_{y}\nonumber \\
 & +\frac{\mu_{c}^{R,S}\mathcal{E}_{31}(t)}{2\sqrt{3}\hbar}L_{z}-\frac{\delta}{2\hbar}(L_{+}^{2}+L_{-}^{2})\label{eq:BHZ_Hamiltonian_time}
\end{align}
where $L_{x}=\frac{\hbar}{\sqrt{2}}\begin{pmatrix}0 & 1 & 0\\
1 & 0 & 1\\
0 & 1 & 0
\end{pmatrix}$, $L_{y}=\frac{\hbar}{\sqrt{2}}\begin{pmatrix}0 & -i & 0\\
i & 0 & -i\\
0 & i & 0
\end{pmatrix}$, $L_{z}=\hbar\begin{pmatrix}1 & 0 & 0\\
0 & 0 & 0\\
0 & 0 & -1
\end{pmatrix}$ are the angular momentum operators for a spin-1 particle and $L_{+}=\sqrt{2}\hbar\begin{pmatrix}0 & 1 & 0\\
0 & 0 & 1\\
0 & 0 & 0
\end{pmatrix}$, $L_{-}$$=\sqrt{2}\hbar\begin{pmatrix}0 & 0 & 0\\
1 & 0 & 0\\
0 & 1 & 0
\end{pmatrix}$ are the corresponding ladder operators. We use the form of the effective
Hamiltonian in Eq. \ref{eq:BHZ_Hamiltonian_time} to calculate the
toplogy of the system. Hereafter, we will assume that the slowly-modulated
electric field amplitudes are
\begin{align}
\mathcal{E}_{21}(t) & =E_{21}\sin(\omega_{1}t),\nonumber \\
\mathcal{E}_{32}(t) & =E_{32}\sin(\omega_{2}t),\nonumber \\
\mathcal{E}_{31}(t) & =E_{31}[m-\cos(\omega_{1}t)-\cos(\omega_{2}t)],\label{eq:BHZ_field}
\end{align}
where $\omega_{1}$, $\omega_{2}$ are two modulation frequencies,
and $m$ is a scalar that characterizes a non-modulated component
of the electric field. These functional forms are inspired from the
TFC scheme reported in Ref. \citen{ boyers2020exploring}.

\noindent \textit{TFC}. ---For completeness, we briefly rederive
the TFC formalism using adiabatic perturbation theory (the original
paper does so within Floquet theory \citep{martin2017topological}).
In the rotating frame, the rate of the system's energy absorption
for the enantiomers is given by $\partial_{t}E^{R,S}(t)=\langle\tilde{\psi}^{R,S}(t)|\partial_{t}\mathcal{H}^{R,S}(t)|\tilde{\psi}^{R,S}(t)\rangle$.
In the long time limit, $t\to\infty,$ the time-averaged energy-absorption
rate, or average power, is \begin{subequations}\label{eq:p_average}
\begin{align}
\mathcal{P}_{av}^{R,S} & =\lim_{t\to\infty}\frac{1}{t}\int_{0}^{t}dt'\partial_{t'}E^{R,S}(t')=\sum_{\omega_{i}}\mathcal{P}_{av}^{R,S}(\omega_{i}),\label{eq:total_pav}\\
\mathcal{P}_{av}^{R,S}(\omega_{i}) & =\lim_{t\to\infty}\frac{1}{t}\int_{0}^{t}dt'\omega_{i}\langle\partial_{\omega_{i}t'}\mathcal{H}^{R,S}(t')\rangle,\label{eq:pav_omega}
\end{align}
\end{subequations}where $\mathcal{P}_{av}^{R,S}(\omega_{i})$ is
the average power at the modulation frequency $\omega_{i}$. 

\begin{figure}
\includegraphics{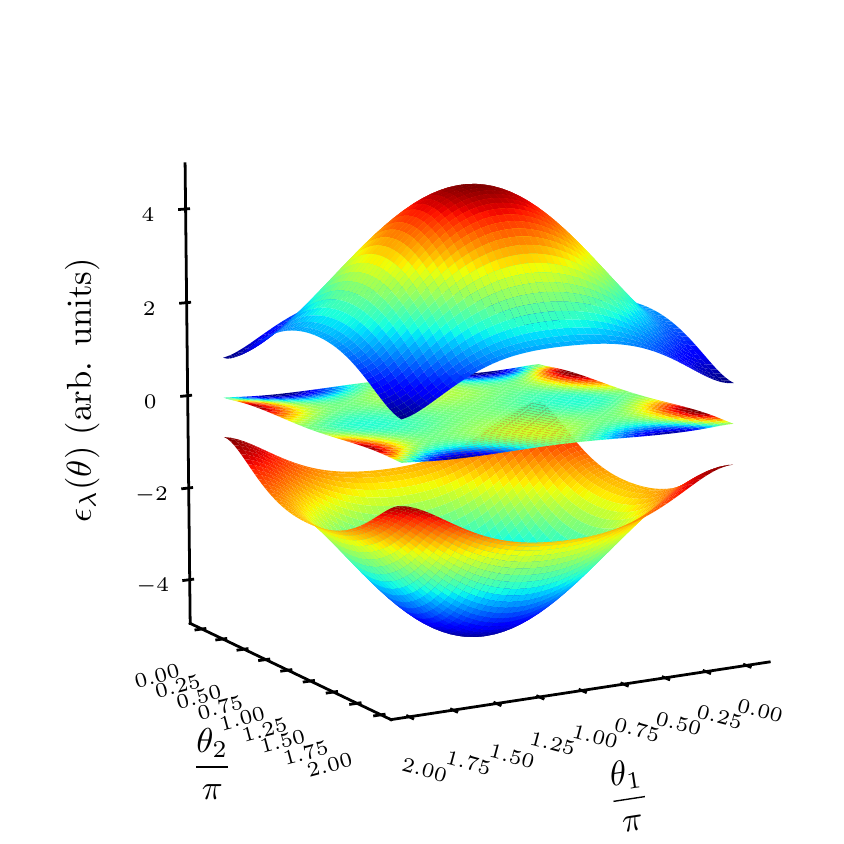}\caption{\emph{\small{}Example adiabatic bands giving rise to enantioselective
TFC.}{\small{} The color gradient is a visual aid for the band dispersion.}\label{fig:Band-structure}}

\end{figure}

Let $|\epsilon_{l}^{R,S}(t)\rangle$ denote the $l-$th adiabatic
state of $\mathcal{H}^{R,S}(t),$ where $\mathcal{H}^{R,S}(t)|\epsilon_{l}^{R,S}(t)\rangle=\epsilon_{l}^{R,S}(t)|\epsilon_{l}^{R,S}(t)\rangle$
(Fig. \ref{fig:Band-structure}). If $\omega_{1},\omega_{2}$ are
incommensurate, \emph{i.e.}, $\omega_{1}/\omega_{2}$ is irrational,
$\mathcal{H}^{R,S}(t)$ is not periodic. However, if we write $\mathcal{H}^{R,S}(t)=\mathcal{H}^{R,S}(\boldsymbol{\theta})=\mathcal{H}^{R,S}(\theta_{1},\theta_{2})$
with $\theta_{i}=\omega_{i}t\,(\text{mod}\,2\pi)$, we notice that
$\mathcal{H}^{R,S}(\boldsymbol{\theta})$ is quasiperiodic, $\text{\ensuremath{\mathcal{\mathcal{H}}^{R,S}(\theta_{1}+2\pi,\theta_{2})=\mathcal{H}^{R,S}(\theta_{1},\theta_{2}+2\pi)=\mathcal{H}^{R,S}(\theta_{1},\theta_{2})}}$,
and the domain of $\mathcal{H}^{R,S}(\theta_{1},\theta_{2})$ is a
two-dimensional torus $\mathbb{T}=[0,2\pi)\otimes[0,2\pi)$.Near the
adiabatic limit where $\omega_{1}$, $\omega_{2}$ are much smaller
than the instantaneous energy gap of $\mathcal{H}^{R,S}(t)$, and
if the system is initiated in the $l-$th adiabatic state, \emph{i.e.,}
$|\tilde{\psi}^{R,S}(0)\rangle=|\epsilon_{l}^{R,S}(0)\rangle$, the
expected quantities $\langle\partial_{\omega_{1}t}\mathcal{H}^{R,S}(t)\rangle$
and $\langle\partial_{\omega_{2}t}\mathcal{H}^{R,S}(t)\rangle$ for
$|\tilde{\psi}^{R,S}(t)\rangle$, to first order in $\omega_{1}$,
$\omega_{2}$ are 

\begin{subequations}
\begin{align}
\langle\partial_{\omega_{1}t}\mathcal{H}^{R,S}(t)\rangle=\langle\partial_{\theta_{1}}\mathcal{H}^{R,S}(\boldsymbol{\theta})\rangle & =\partial_{\theta_{1}}\epsilon_{l}^{R,S}(\boldsymbol{\theta})-\hbar\omega_{2}F_{l}^{R,S}(\boldsymbol{\theta})\label{eq:photon_rate_1}\\
\langle\partial_{\omega_{2}t}\mathcal{H}^{R,S}(t)\rangle=\langle\partial_{\theta_{2}}\mathcal{H}^{R,S}(\boldsymbol{\theta})\rangle & =\partial_{\theta_{2}}\epsilon_{l}^{R,S}(\boldsymbol{\theta})+\hbar\omega_{1}F_{l}^{R,S}(\boldsymbol{\theta})\label{eq:photon_rate_2}
\end{align}
\end{subequations}where $F_{l}^{R,S}(\boldsymbol{\theta})=i\langle\partial_{\theta_{1}}\epsilon_{l}^{R,S}(\boldsymbol{\theta})|\partial_{\theta_{2}}\epsilon_{l}^{R,S}(\boldsymbol{\theta})\rangle+\text{h.c.}$
is the Berry curvature of the $l-$th adiabatic band (see, SI-3).

According to the mean-value theorem for incommensurate $\omega_{1}$,
$\omega_{2}$ \citep{samoilenko2012elements}, the linear flow of
$\boldsymbol{\theta}$ covers the torus densely for long enough times.
Thus, the time average of $F_{l}^{R,S}(t)$ is the same as the average
of $F_{l}^{R,S}(\boldsymbol{\theta})$ over the entire torus $\mathbb{T}$:
\begin{equation}
\lim_{t\to\infty}\frac{1}{t}\int_{0}^{t}dt'F_{l}^{R,S}(t')=\frac{1}{4\pi^{2}}\int_{\mathbb{T}}d\boldsymbol{\theta}F_{l}^{R,S}(\boldsymbol{\theta}).\label{eq:Mean_value_theorem}
\end{equation}

From a practical standpoint, $t\to\infty$ means $t>p\frac{2\pi}{\omega_{1}}=q\frac{2\pi}{\omega_{2}}$,
where $\frac{\omega_{1}}{\omega_{2}}\approx\frac{p}{q}$ for $p,q\in\mathbb{Z}^{+}$.
Substituting Eqs. \ref{eq:photon_rate_1}-\ref{eq:photon_rate_2}
into Eq. \ref{eq:pav_omega} gives rise to the average power lost
by the fields at $\omega_{1},$ $\omega_{2}$ when the system is initiated
in the $l-$th band, $\mathcal{P}_{av}^{R,S}(\omega_{1})=-\mathcal{P}_{av}^{R,S}(\omega_{2})=-\frac{\hbar\omega_{1}\omega_{2}C_{l}^{R,S}}{2\pi}$.
Here the average of $\partial_{\theta_{i}}\epsilon_{l}^{R,S}(\boldsymbol{\theta})$
is zero since $\epsilon_{l}^{R,S}(\boldsymbol{\theta})$ is quasiperiodic
in $\boldsymbol{\theta}$, and $C_{l}^{R,S}=\frac{1}{2\pi}\int_{\mathbb{T}}d\boldsymbol{\theta}F_{l}^{R,S}(\boldsymbol{\theta})$
is the Chern number of the $l-$th band for the corresponding enantiomer.
Thus, the enantiomer dependent average energy-pumping rate between
the two modulation fields $\mathcal{P}_{2\to1}^{R,S}=[\mathcal{P}_{av}^{R,S}(\omega_{2})-\mathcal{P}_{av}^{R,S}(\omega_{1})]/2$
is quantized,
\begin{align}
\mathcal{P}_{2\to1}^{R,S} & =\frac{\hbar\omega_{1}\omega_{2}C_{l}^{R,S}}{2\pi},\label{eq:pumping_rate}
\end{align}
or in other words, after one period of the $\omega_{2}$ modulation,
$C_{l}^{R,S}$ photons with frequency $\omega_{1}$ are produced.
The photons produced are in the same spatial modes as the incoming
electric fields. The very off-resonant nature of this process guarantees
that the molecule does not retain energy and the energy transfer process
occurs only between the fields.

\begin{figure}
\includegraphics{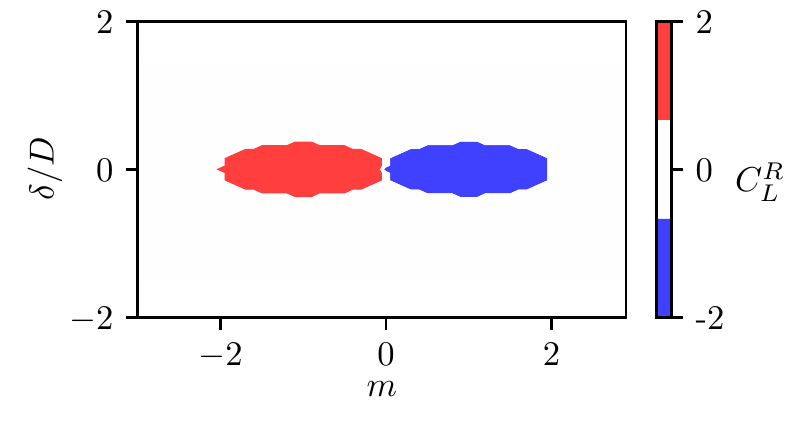}\caption{\emph{\small{}Topological phase diagram. }{\small{}The value of $C_{L}^{R}$
is calculated taking the magnitudes of the light-matter couplings
to be equal, }\emph{\small{}i.e., }{\small{}$|\mu_{2,M';1M}^{R}E_{21}|=|\mu_{3,M';2,M}^{R}E_{32}|=|\mu_{3,M';1,M}^{R}E_{31}|=\hbar D$,
while the laser-driving parameters $m$ and $\delta$ are varied.
We obtain $C_{L}^{R}=-2\text{sgn}(m)\text{sgn}(\mathcal{O}^{R})$
at the vicinity of $\delta=0$, where $\mathcal{O}^{R}=-\mathcal{O}^{S}$
is the Král-Shapiro product, which is enantioselective.} \label{fig:Topological-phase-diagram.}}
\end{figure}

For $\delta=0,$ $\mathcal{H}^{R,S}(\boldsymbol{\theta})$ (see Eq.
\ref{eq:BHZ_Hamiltonian_time}), resembles half of the Bernevig-Hughes-Zhang
Hamiltonian \citep{bernevig2006quantum}, except that the Pauli matrices
are replaced with the spin-1 angular momentum operators. As expected,
$\mathcal{H}^{R,S}(\boldsymbol{\theta})$ is topologically non-trivial
for $|m|<2$, where the Chern numbers for the upper ($U$) and lower
($L$) adiabatic bands remarkably acquire the value,
\begin{equation}
C_{U}^{R,S}=2\text{sgn}(m)\text{sgn}(\mathcal{O}^{R,S})=-C_{L}^{R,S},\label{eq:Chern_number}
\end{equation}
and that for the middle ($M$) band $C_{M}=0$ (for an analytical
proof, see SI-4 ). $\mathcal{O}^{R,S}=(\mu_{b}^{R,S}E_{12})(\mu_{a}^{R,S}E_{23})(\mu_{c}^{R,S}E_{31})$
is the KS product which obeys the enantioselective symmetry relation
$\mathcal{O}^{R}=-\mathcal{O}^{S}$, since $\mu_{a}^{R}\mu_{b}^{R}\mu_{c}^{R}=-\mu_{a}^{S}\mu_{b}^{S}\mu_{c}^{S}$,
and we have assumed that $E_{ij}=E_{ji}$. Therefore $C_{L}^{R}=-C_{L}^{S}$,
and the TFC for the two enantiomers initiated in the lower (upper)
adiabatic band at $t=0$ is expected to have the same magnitude but
opposite sign, \emph{i.e.}, $\mathcal{P}_{2\to1}^{R}=-\mathcal{P}_{2\to1}^{S}$.
This results begs us to consider the fruitful analogy between enantiomer
label and spin degrees of freedom. Just like in the QSHE, where the
transverse conductivity for opposite spins bears opposite signs, so
does the TFC for opposite enantiomers. Eq. \ref{eq:Chern_number}
is the central result of this letter and relates a fundamental topological
invariant from chiroptical spectroscopy ($\text{sgn}\mathcal{O}=\pm1$)
with the notions of SPTPs. Fig. \ref{fig:Topological-phase-diagram.}
shows the computed value of $C_{L}^{R}$ for different values of $m$
when $\delta\neq0.$

By analogy with Eq. \ref{eq:p_average}, we can compute the enantiomer
dependent average power absorbed in the original frame as $P_{av}^{R,S}(\Omega)=\lim_{t\to\infty}\frac{1}{t}\int_{0}^{t}dt'\Omega\langle\partial_{\Omega t'}H^{R,S}(t')\rangle$,
obtaining:\begin{small}\begin{subequations}
\begin{align}
\frac{\mathcal{P}_{av}^{R,S}(\omega_{1})}{\hbar\omega_{1}} & =\frac{P_{av}^{R,S}(\Omega_{21+1})}{\hbar\Omega_{21+1}}-\frac{P_{av}^{R,S}(\Omega_{21-1})}{\hbar\Omega_{21-1}}\label{eq:n_1_pumping_rate}\\
 & +\frac{P_{av}^{R,S}(\Omega_{31+1})}{\hbar\Omega_{31+1}}-\frac{P_{av}^{R,S}(\Omega_{31-1})}{\hbar\Omega_{31-1}},\nonumber \\
\frac{\mathcal{P}_{av}^{R,S}(\omega_{2})}{\hbar\omega_{1}} & =\frac{P_{av}^{R,S}(\Omega_{32+2})}{\hbar\Omega_{32+2}}-\frac{P_{av}^{R,S}(\Omega_{32-2})}{\hbar\Omega_{32-2}}\nonumber \\
 & +\frac{P_{av}^{R,S}(\Omega_{31+2})}{\hbar\Omega_{31+2}}-\frac{P_{av}^{R,S}(\Omega_{31-2})}{\hbar\Omega_{31-2}},\label{eq:n_2_pumping_rate}
\end{align}
\end{subequations}\end{small} where $\Omega_{ij\pm1,2}=\Omega_{ij}\pm\omega_{1,2}$.
Thus, the quantization due to the enantioselective TFC can be extracted
from an experimentally detected difference power spectrum of the fields
interacting with the molecule. Notice that the topology is preserved
for $\delta\neq0$ as long as $\hbar|\delta|<|\mu_{i,M';j,M}^{R,S}E_{ij}|/2$.
In general, we expect our scheme to maintain the nontrivial topology
with respect to changes in experimental conditions (such as laser
spot size or collection efficiency) so long as adiabaticity still
holds and the necessary peaks in the power spectrum can be resolved.

\begin{figure}[h]
\includegraphics{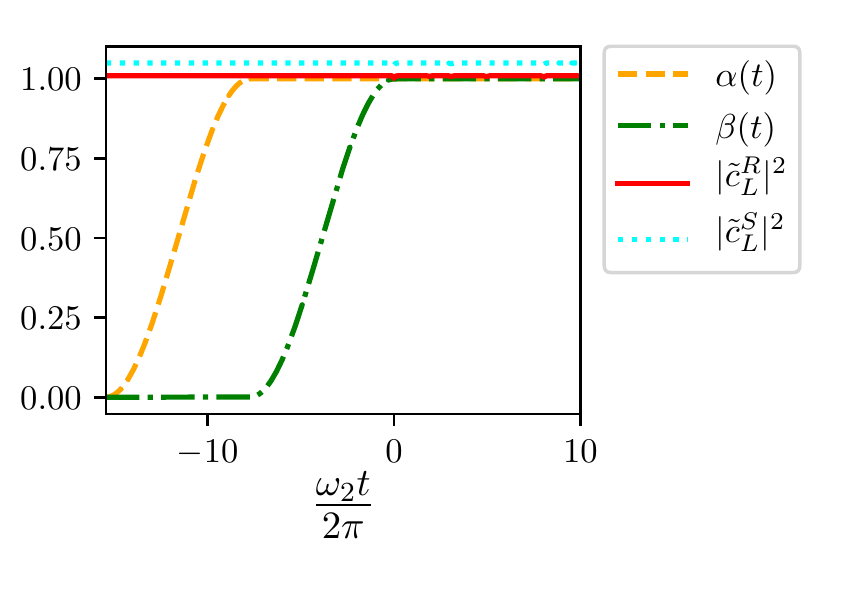}

\caption{\emph{\small{}Adiabatic state preparation.}{\small{} Presented are
the plots for the functions $\alpha(t)$ and $\beta(t)$. We also
feature the populations $|\tilde{c}_{L}^{R}|^{2}$, $|\tilde{c}_{L}^{S}|^{2}$
(shifted vertically slightly to be visible) of the lower adiabatic
state for each enantiomer. As shown, the system is effectively prepared
in the lower adiabatic bands for both enantiomers. }\label{fig:adiabaticity}}
\end{figure}

The dynamics of the system is calculated by numerically integrating
the Schrödinger equation in the rotating frame (Eq. \ref{eq:rotating_frame_hamiltonian}),
and the power spectrum is obtained by returning to the original frame.
In atomic units ($\hbar=1$), the electric field amplitudes are taken
to be $E_{21}=5\sqrt{3}E_{0}$, $E_{32}=6E$, $E_{31}=\sqrt{3}E_{0}$,
where $E_{0}=4.0\times10^{-9}\text{\;a.u.}$, the dipole moment principal
axes components are $\mu_{a}^{R}=\text{\ensuremath{\mu_{a}^{S}}}=0.47\text{\;a.u.}$,
$\mu_{b}^{R}=\ensuremath{\mu_{b}^{S}}=0.75\text{\;a.u.}$, $\mu_{c}^{R}=-\text{\ensuremath{\mu_{c}^{S}}}=0.14\text{\;a.u.}$,
and the molecular transition energies are $\epsilon_{2}-\epsilon_{1}=4.4\times10^{-8}\;\text{a.u.}$
and $\epsilon_{3}-\epsilon_{1}=4.7\times10^{-8}\;\text{a.u.}$ The
dipole moment components and molecular energies are extracted from
a microwave three-wave-mixing model for $R-$ and $S-$1,2-propanediol
\citep{patterson2013enantiomer}. Using these parameters, it is true
that $|\mu_{i,M';j,M}^{R,S}E_{ij}|/2\ll\hbar\Omega_{ij}$, so the
rotating wave approximation holds. The slow incommensurate modulation
frequencies and laser detuning are taken to be $\omega_{1}=\omega_{2}/\phi=\delta=1\times10^{-11}\text{\;a.u.}$,
where we take $\phi=\frac{\sqrt{5}-1}{2}$ as in Ref.  \citen{martin2017topological},
satisfying the perturbative condition $\hbar|\delta|,\hbar\omega_{1},\hbar\omega_{2}\ll|\mu_{i,M;j,M'}^{R,S}\cdot E_{ij}|/2$.
Setting $m=1.4$, the system is in the topologically nontrivial regime.

\begin{figure}[h]
\includegraphics{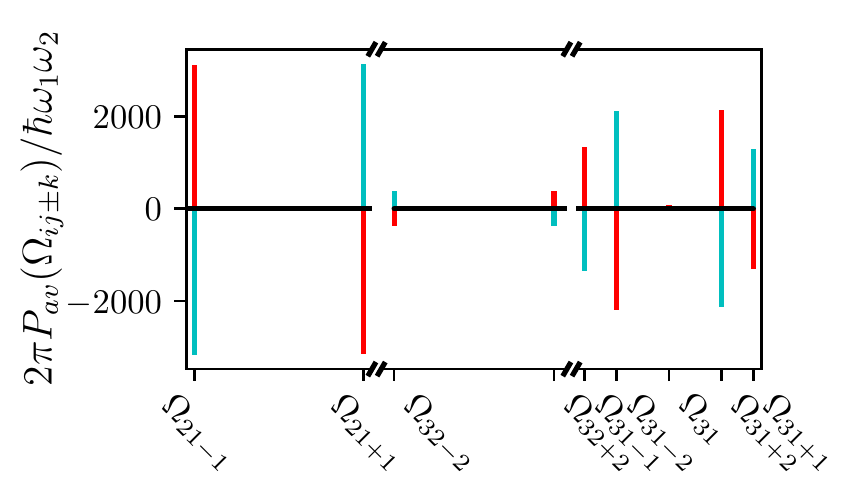}

\caption{\emph{\small{}Enantioselective TFC. }{\small{}Plotted is the difference
power spectrum for the driving electric field when coupled to a single
$R-$ (red) and $S-$ (cyan) 1,2-propanediol enantiomers. In terms
of intensity, assuming the laser beam waist area is $\sim1$ $\text{cm}^{2}$,
the change observed is $\sim10^{-15}\,\text{W}\cdot\text{m}^{-2}$
per molecule. This spectrum is enantioselective, and using Eqs. \ref{eq:n_1_pumping_rate}
and \ref{eq:n_2_pumping_rate}, we can see that the frequency conversion
in the rotating frame is topological, $\frac{2\pi}{\hbar\omega_{1}\omega_{2}}\mathcal{P}_{2\to1}^{R}=-\frac{2\pi}{\hbar\omega_{1}\omega_{2}}\mathcal{P}_{1\to2}^{S}=-2$.}
\label{fig:powerspectrum}}
\end{figure}

To obtain the desired enantioselective TFC, both enantiomers need
to be prepared in the lowest adiabatic bands in the rotating frame
at $t=0$. Suppose that before fields are turned on ($\mu_{i,M';j,M}^{R,S}\mathcal{E}_{ij}(t)\to0$
as $t\to-\infty$), the molecules start in the ground state $|1,0\rangle$.
Under those circumstances, the eigenstates of Eq. \ref{eq:rotating_frame_hamiltonian}
are the states $|1,0\rangle,|2,M\rangle,|3,0\rangle$ with eigenenergies
$\epsilon_{L,M,U}^{R,S}(-\infty)=-\delta,0,\delta$, and the state
of each molecule is $|\epsilon_{L}^{R,S}(-\infty)\rangle$. If the
electric fields are slowly turned on at a rate $\omega_{r}$ that
is much smaller than the instantaneous band gaps $|\epsilon_{l}^{R,S}(t)-\epsilon_{l'}^{R,S}(t)|$,
both enantiomers are prepared in the lower band, \emph{i.e.} $|\epsilon_{L}^{R,S}(0)\rangle$.
Note that the modulating frequencies $\omega_{1},\:\omega_{2}$ must
also be much smaller than $|\epsilon_{l}^{R,S}(t)-\epsilon_{l'}^{R,S}(t)|$
at all times. Chirped microwave fields for $t<0$ satisfy this constraint.
The adiabatic protocol we choose is $E_{ij}\to E_{ij}\alpha(t)$ and
$\omega_{1,2}\to\omega_{1,2}\beta(t)$, where the ramp-up functions
slowly vary at the rate $\omega_{r}=2\times10^{-13}\text{\;a.u.}$
(see Fig. \ref{fig:adiabaticity}),

\begin{small}\begin{subequations}

\begin{align}
\alpha(t)= & \begin{cases}
0 & t<-\frac{2\pi}{\omega_{r}},\\
\frac{1-\text{cos}\omega_{r}t}{2} & -\frac{2\pi}{\omega_{r}}<t<-\frac{\pi}{\omega_{r}},\\
1 & -\frac{\pi}{\omega_{r}}<t,
\end{cases}\label{eq:alpha}\\
\beta(t)= & \begin{cases}
0 & t<-\frac{\pi}{\omega_{r}},\\
\frac{1+\text{cos}\omega_{r}t}{2} & -\frac{\pi}{\omega_{r}}<t<0,\\
1 & 0<t.
\end{cases}\label{eq:beta}
\end{align}

\end{subequations}\end{small} After a sufficiently long time (we
choose $t^{*}=2000\times2\pi/\omega_{2}$), the frequency-resolved
time-averaged power spectrum $P_{av}(\Omega)$ lost by the fields
is numerically calculated considering only $t\geq0$. This quantity
is indeed enantioselective, and using Eqs. \ref{eq:n_1_pumping_rate}
and \ref{eq:n_2_pumping_rate}, each enantiomer Chern number for the
lower band $C_{L}^{R}=-2=-C_{L}^{S}$ is extracted, revealing the
topological nature of this nonlinear optical phenomenon (Fig. \ref{fig:powerspectrum}).

For an ensemble containing $N_{R}$ $R-$molecules and $N_{S}$ $S-$molecules,
which are all prepared in the ground state $|1,0\rangle$, the expected
pumping rate is 
\begin{equation}
\mathcal{P}_{2\to1}=\frac{\hbar\omega_{1}\omega_{2}C_{L}^{R}}{2\pi}(N_{R}-N_{S}).\label{eq:ensemble_pumping_rate}
\end{equation}
which is zero for a racemic mixture, but otherwise, reveals the EE
$|N_{R}-N_{S}|$ and chirality $\text{sgn}(N_{R}-N_{S})$. Notice
that in line with other nonlinear chiroptical signals that depend
on electric but not magnetic dipole contributions \citep{ordonez2018generalized},
Eq. \ref{eq:ensemble_pumping_rate} contains no background achiral
signal, unlike traditional circular dichroism, where both enantiomers
have the same electric dipole and magnetic dipole absorption strengths
for circularly polarized light \citep{schellman1975circular}.

Let us briefly discuss the limits of enantioselective TFC. First,
the linewidths of microwave transitions are on the order of 10-100
kHz \citep{park2016perspective}, which are smaller than the adiabatic
state preparation gap $\delta\approx1\,\text{MHz}$, as well as the
light-matter interactions $|\mu_{ij}E_{ij}|/\hbar\approx10\,\text{MHz}$
inducing the topological gap, or even the smallest difference in energies
in the power spectrum (see, Fig. \ref{fig:powerspectrum}, $\Omega_{31\pm1}-\Omega_{31\pm2}\approx1\,\text{MHz}$).
Thus, the described protocol should be resilient to the finite linewidths
of these transitions. Second, another source of imperfections stems
from laser shot noise. Assuming that the laser beam waist area is
$\sim1\:\text{cm}^{2}$ and considering the field strength above,
the shot noise for a time interval $t^{*}$ is $\sqrt{N}\sim10^{9}$
(where $N$ is the expected number of photons produced by the field,
see SI-5). From the power spectrum (Fig. \ref{fig:powerspectrum}),
we find that for the same time interval, that the minimal magnitude
of the change in the photon number due to the TFC is $\min\Big(\left|\frac{P_{av}(\Omega_{ij\pm1,2})t^{*}}{\hbar\Omega_{ij\pm1,2}}\right|\Big)\approx100\times|N_{R}-N_{S}|.$
Therefore, as long as the magnitude of the enantiomer excess $|N_{R}-N_{S}|$
is much larger than $\sim10^{7}$ molecules, the signal should be
detectable above the shot noise. In terms of percentage of the total
molecule count $N_{R}+N_{S}$, the lower end of the EE detection limit
for $1$ mL of a $1$ $\mu$M solution is $10^{-6}\%$. We conclude
with a few comments on the observability of our predictions. First,
this study has assumed the ideal limit that the molecules are at $0$
K. Under typical experimental conditions for microwave-three wave
mixing at 7 K \citep{patterson2013enantiomer} all three rotational
energy levels used in our model are substantially thermally occupied.
In this scenario, enantioselective frequency conversion still survives;
however, the integer Chern number will be replaced by a thermal average
of the Chern numbers $C_{L}^{R},C_{M}^{R},C_{U}^{R}$. Second, the
excited-state thermal populations can be bypassed by working in a
different energy range, such as the UV-visible one involving electronic
transitions and the infrared one involving vibrations \citep{belkin2000sum,fischer2003new};
the price to pay in the first case is the complication introduced
by electron-vibration coupling. These complications will be addressed
in future works.

In summary, we have presented an enantioselective TFC setup for an
ensemble of chiral molecules. Owing to the dependence of the topological
invariant on the sign of the KS product (Eq. \ref{eq:Chern_number}),
which differs by a phase of $\pi$ for the two enantiomers, the quantized
time-averaged energy-pumping rate is of opposite sign for the $R-$
and $S-$ molecules, just like transverse conductivity is of opposite
sign for up and down spins in the QSHE. We show that the computed
signal is non-zero for any sample with EE and vanishes for a racemic
mixture. An intriguing consequence of Eq. \ref{eq:pumping_rate} is
that as long as the timescale separations required by the model are
fulfilled, the chemical identity of the probed molecules (\emph{e.g.},
through the strengths of the transition dipole moments) in the rotating
frame is erased by the signal, leading to a universal nonlinear optical
response which acknowledges the enantiomeric excess only. This characteristic
is reminiscent to the very accurate determination of the quantum of
conductance with a wide range of QHE systems. Thus, from a metrological
standpoint, the generality of the enantioselective TFC can be exploited
to accurately measure EE by running a linear fit of the pumping rate
$\mathcal{P}_{2\rightarrow1}$ for a series of experiments where $\omega_{1}$
(or $\omega_{2}$) is varied. Furthermore, if one is only concerned
with $|\text{EE}|$, a practical asset of the presented methodology
is that there is no need to calibrate the signal with an enantiopure
sample beforehand. We believe that the removal of calibration counterbalances
the complexity of the experimental setup proposed in this Letter. 

While concepts of topology have been very productive in the exploration
of new condensed matter physics phenomena, most of them are restricted
to periodic solids (see Ref. \citen{faure2000topological,schwennicke2020optical}
for a few molecular exceptions). TFC \citep{martin2017topological,nathan2019topological}
is a powerful tool that opens doors to the application of those concepts
to 0D systems such as finite molecular and non-periodic nanoscale
systems. In particular, this work reveals that laser-dressed chiral
molecules support SPTPs that are not adiabatically connected to their
non-laser-dressed counterparts. It also provides a fruitful playground
to explore further conceptual connections between topological physics
and molecular chirality \citealp{ordonez2019propensity,ordonez2021geometric}

\section*{Acknowledgments}

K.S. and J.Y.- Z. were supported by NSF CAREER CHE 1654732. This work
employed computational resources of the Extreme Science and Engineering
Discovery Environment (XSEDE), which is supported by National Science
Foundation Grant No. ACI-1548562, under allocation No. TG-ASC150024.
K.S. acknowledges helpful discussions with Matthew Du, Stephan van
den Wildenberg, and Jorge A. Campos-González-Angulo. Both K.S. and
J.Y.-Z. acknowledge the help of insightful comments from previous
Reviewers. 

\section*{Supporting Information Available: }

Presentation of rotational eigenstates of an asymmetric top; discussion
of the change of basis applied to obtain the necessary effective Hamiltonian,
review of adiabatic perturbation theory; derivation of the expected
laser power in the rotating frame; analytical evaluation of enantioselective
Chern number for zero detuning; calculation of expected laser shot
noise. 

\bibliography{manuscript_03_04_22}

\providecommand{\latin}[1]{#1}
\makeatletter
\providecommand{\doi}
  {\begingroup\let\do\@makeother\dospecials
  \catcode`\{=1 \catcode`\}=2 \doi@aux}
\providecommand{\doi@aux}[1]{\endgroup\texttt{#1}}
\makeatother
\providecommand*\mcitethebibliography{\thebibliography}
\csname @ifundefined\endcsname{endmcitethebibliography}
  {\let\endmcitethebibliography\endthebibliography}{}
\begin{mcitethebibliography}{63}
\providecommand*\natexlab[1]{#1}
\providecommand*\mciteSetBstSublistMode[1]{}
\providecommand*\mciteSetBstMaxWidthForm[2]{}
\providecommand*\mciteBstWouldAddEndPuncttrue
  {\def\EndOfBibitem{\unskip.}}
\providecommand*\mciteBstWouldAddEndPunctfalse
  {\let\EndOfBibitem\relax}
\providecommand*\mciteSetBstMidEndSepPunct[3]{}
\providecommand*\mciteSetBstSublistLabelBeginEnd[3]{}
\providecommand*\EndOfBibitem{}
\mciteSetBstSublistMode{f}
\mciteSetBstMaxWidthForm{subitem}{(\alph{mcitesubitemcount})}
\mciteSetBstSublistLabelBeginEnd
  {\mcitemaxwidthsubitemform\space}
  {\relax}
  {\relax}

\bibitem[Pasteur(1848)]{pasteur1848relations}
Pasteur,~L. Sur les relations qui peuvent exister entre la forme crystalline,
  la composition chimique et le sens de la polarization rotatoire. \emph{Ann.
  Chim. Phys.} \textbf{1848}, \emph{24}, 442--459\relax
\mciteBstWouldAddEndPuncttrue
\mciteSetBstMidEndSepPunct{\mcitedefaultmidpunct}
{\mcitedefaultendpunct}{\mcitedefaultseppunct}\relax
\EndOfBibitem
\bibitem[Quack \latin{et~al.}(2008)Quack, Stohner, and Willeke]{quack2008high}
Quack,~M.; Stohner,~J.; Willeke,~M. High-resolution spectroscopic studies and
  theory of parity violation in chiral molecules. \emph{Annu. Rev. Phys. Chem.}
  \textbf{2008}, \emph{59}, 741--769\relax
\mciteBstWouldAddEndPuncttrue
\mciteSetBstMidEndSepPunct{\mcitedefaultmidpunct}
{\mcitedefaultendpunct}{\mcitedefaultseppunct}\relax
\EndOfBibitem
\bibitem[Hutt and Tan(1996)Hutt, and Tan]{hutt1996drug}
Hutt,~A.~J.; Tan,~S.~C. Drug chirality and its clinical significance.
  \emph{Drugs} \textbf{1996}, \emph{52}, 1--12\relax
\mciteBstWouldAddEndPuncttrue
\mciteSetBstMidEndSepPunct{\mcitedefaultmidpunct}
{\mcitedefaultendpunct}{\mcitedefaultseppunct}\relax
\EndOfBibitem
\bibitem[Kasprzyk-Hordern(2010)]{kasprzyk2010pharmacologically}
Kasprzyk-Hordern,~B. Pharmacologically active compounds in the environment and
  their chirality. \emph{Chem. Soc. Rev.} \textbf{2010}, \emph{39},
  4466--4503\relax
\mciteBstWouldAddEndPuncttrue
\mciteSetBstMidEndSepPunct{\mcitedefaultmidpunct}
{\mcitedefaultendpunct}{\mcitedefaultseppunct}\relax
\EndOfBibitem
\bibitem[Blackmond(2010)]{blackmond2010origin}
Blackmond,~D.~G. The origin of biological homochirality. \emph{Cold Spring
  Harbor Perspect. Biol.} \textbf{2010}, \emph{2}, a002147\relax
\mciteBstWouldAddEndPuncttrue
\mciteSetBstMidEndSepPunct{\mcitedefaultmidpunct}
{\mcitedefaultendpunct}{\mcitedefaultseppunct}\relax
\EndOfBibitem
\bibitem[Tang and Cohen(2010)Tang, and Cohen]{tang2010optical}
Tang,~Y.; Cohen,~A.~E. Optical chirality and its interaction with matter.
  \emph{Phys. Rev. Lett.} \textbf{2010}, \emph{104}, 163901\relax
\mciteBstWouldAddEndPuncttrue
\mciteSetBstMidEndSepPunct{\mcitedefaultmidpunct}
{\mcitedefaultendpunct}{\mcitedefaultseppunct}\relax
\EndOfBibitem
\bibitem[Poulikakos \latin{et~al.}(2018)Poulikakos, Thureja, Stollmann, De~Leo,
  and Norris]{poulikakos2018chiral}
Poulikakos,~L.~V.; Thureja,~P.; Stollmann,~A.; De~Leo,~E.; Norris,~D.~J. Chiral
  light design and detection inspired by optical antenna theory. \emph{Nano
  Lett.} \textbf{2018}, \emph{18}, 4633--4640\relax
\mciteBstWouldAddEndPuncttrue
\mciteSetBstMidEndSepPunct{\mcitedefaultmidpunct}
{\mcitedefaultendpunct}{\mcitedefaultseppunct}\relax
\EndOfBibitem
\bibitem[Garc{\'\i}a-Guirado \latin{et~al.}(2018)Garc{\'\i}a-Guirado,
  Svedendahl, Puigdollers, and Quidant]{garcia2018enantiomer}
Garc{\'\i}a-Guirado,~J.; Svedendahl,~M.; Puigdollers,~J.; Quidant,~R.
  Enantiomer-selective molecular sensing using racemic nanoplasmonic arrays.
  \emph{Nano Lett.} \textbf{2018}, \emph{18}, 6279--6285\relax
\mciteBstWouldAddEndPuncttrue
\mciteSetBstMidEndSepPunct{\mcitedefaultmidpunct}
{\mcitedefaultendpunct}{\mcitedefaultseppunct}\relax
\EndOfBibitem
\bibitem[Ordonez and Smirnova(2018)Ordonez, and
  Smirnova]{ordonez2018generalized}
Ordonez,~A.~F.; Smirnova,~O. Generalized perspective on chiral measurements
  without magnetic interactions. \emph{Phys. Rev. A} \textbf{2018}, \emph{98},
  063428\relax
\mciteBstWouldAddEndPuncttrue
\mciteSetBstMidEndSepPunct{\mcitedefaultmidpunct}
{\mcitedefaultendpunct}{\mcitedefaultseppunct}\relax
\EndOfBibitem
\bibitem[Ritchie(1976)]{ritchie1976theory}
Ritchie,~B. Theory of the angular distribution of photoelectrons ejected from
  optically active molecules and molecular negative ions. \emph{Phys. Rev. A}
  \textbf{1976}, \emph{13}, 1411\relax
\mciteBstWouldAddEndPuncttrue
\mciteSetBstMidEndSepPunct{\mcitedefaultmidpunct}
{\mcitedefaultendpunct}{\mcitedefaultseppunct}\relax
\EndOfBibitem
\bibitem[B{\"o}wering \latin{et~al.}(2001)B{\"o}wering, Lischke, Schmidtke,
  M{\"u}ller, Khalil, and Heinzmann]{bowering2001asymmetry}
B{\"o}wering,~N.; Lischke,~T.; Schmidtke,~B.; M{\"u}ller,~N.; Khalil,~T.;
  Heinzmann,~U. Asymmetry in photoelectron emission from chiral molecules
  induced by circularly polarized light. \emph{Phys. Rev. Lett.} \textbf{2001},
  \emph{86}, 1187\relax
\mciteBstWouldAddEndPuncttrue
\mciteSetBstMidEndSepPunct{\mcitedefaultmidpunct}
{\mcitedefaultendpunct}{\mcitedefaultseppunct}\relax
\EndOfBibitem
\bibitem[Lux \latin{et~al.}(2012)Lux, Wollenhaupt, Bolze, Liang, K{\"o}hler,
  Sarpe, and Baumert]{lux2012circular}
Lux,~C.; Wollenhaupt,~M.; Bolze,~T.; Liang,~Q.; K{\"o}hler,~J.; Sarpe,~C.;
  Baumert,~T. Circular dichroism in the photoelectron angular distributions of
  camphor and fenchone from multiphoton ionization with femtosecond laser
  pulses. \emph{Angew. Chem., Int. Ed.} \textbf{2012}, \emph{51},
  5001--5005\relax
\mciteBstWouldAddEndPuncttrue
\mciteSetBstMidEndSepPunct{\mcitedefaultmidpunct}
{\mcitedefaultendpunct}{\mcitedefaultseppunct}\relax
\EndOfBibitem
\bibitem[Demekhin \latin{et~al.}(2018)Demekhin, Artemyev, Kastner, and
  Baumert]{demekhin2018photoelectron}
Demekhin,~P.~V.; Artemyev,~A.~N.; Kastner,~A.; Baumert,~T. Photoelectron
  circular dichroism with two overlapping laser pulses of carrier frequencies
  $\omega$ and 2 $\omega$ linearly polarized in two mutually orthogonal
  directions. \emph{Phys. Rev. Lett.} \textbf{2018}, \emph{121}, 253201\relax
\mciteBstWouldAddEndPuncttrue
\mciteSetBstMidEndSepPunct{\mcitedefaultmidpunct}
{\mcitedefaultendpunct}{\mcitedefaultseppunct}\relax
\EndOfBibitem
\bibitem[Fischer \latin{et~al.}(2000)Fischer, Wiersma, Righini, Champagne, and
  Buckingham]{fischer2000three}
Fischer,~P.; Wiersma,~D.~S.; Righini,~R.; Champagne,~B.; Buckingham,~A.~D.
  Three-wave mixing in chiral liquids. \emph{Phys. Rev. Lett.} \textbf{2000},
  \emph{85}, 4253\relax
\mciteBstWouldAddEndPuncttrue
\mciteSetBstMidEndSepPunct{\mcitedefaultmidpunct}
{\mcitedefaultendpunct}{\mcitedefaultseppunct}\relax
\EndOfBibitem
\bibitem[Brumer \latin{et~al.}(2001)Brumer, Frishman, and
  Shapiro]{brumer2001principles}
Brumer,~P.; Frishman,~E.; Shapiro,~M. Principles of electric-dipole-allowed
  optical control of molecular chirality. \emph{Phys. Rev. A} \textbf{2001},
  \emph{65}, 015401\relax
\mciteBstWouldAddEndPuncttrue
\mciteSetBstMidEndSepPunct{\mcitedefaultmidpunct}
{\mcitedefaultendpunct}{\mcitedefaultseppunct}\relax
\EndOfBibitem
\bibitem[Beaulieu \latin{et~al.}(2018)Beaulieu, Comby, Descamps, Fabre, Garcia,
  G{\'e}neaux, Harvey, L{\'e}gar{\'e}, Ma{\v{s}}{\'\i}n, Nahon, \latin{et~al.}
  others]{beaulieu2018photoexcitation}
Beaulieu,~S.; Comby,~A.; Descamps,~D.; Fabre,~B.; Garcia,~G.~A.;
  G{\'e}neaux,~R.; Harvey,~A.~G.; L{\'e}gar{\'e},~F.; Ma{\v{s}}{\'\i}n,~Z.;
  Nahon,~L., \latin{et~al.}  Photoexcitation circular dichroism in chiral
  molecules. \emph{Nat. Phys.} \textbf{2018}, \emph{14}, 484--489\relax
\mciteBstWouldAddEndPuncttrue
\mciteSetBstMidEndSepPunct{\mcitedefaultmidpunct}
{\mcitedefaultendpunct}{\mcitedefaultseppunct}\relax
\EndOfBibitem
\bibitem[Neufeld \latin{et~al.}(2019)Neufeld, Ayuso, Decleva, Ivanov, Smirnova,
  and Cohen]{neufeld2019ultrasensitive}
Neufeld,~O.; Ayuso,~D.; Decleva,~P.; Ivanov,~M.~Y.; Smirnova,~O.; Cohen,~O.
  Ultrasensitive chiral spectroscopy by dynamical symmetry breaking in high
  harmonic generation. \emph{Phys. Rev. X} \textbf{2019}, \emph{9},
  031002\relax
\mciteBstWouldAddEndPuncttrue
\mciteSetBstMidEndSepPunct{\mcitedefaultmidpunct}
{\mcitedefaultendpunct}{\mcitedefaultseppunct}\relax
\EndOfBibitem
\bibitem[Ayuso \latin{et~al.}(2019)Ayuso, Neufeld, Ordonez, Decleva, Lerner,
  Cohen, Ivanov, and Smirnova]{ayuso2019synthetic}
Ayuso,~D.; Neufeld,~O.; Ordonez,~A.~F.; Decleva,~P.; Lerner,~G.; Cohen,~O.;
  Ivanov,~M.; Smirnova,~O. Synthetic chiral light for efficient control of
  chiral light--matter interaction. \emph{Nat. Photonics} \textbf{2019},
  \emph{13}, 866--871\relax
\mciteBstWouldAddEndPuncttrue
\mciteSetBstMidEndSepPunct{\mcitedefaultmidpunct}
{\mcitedefaultendpunct}{\mcitedefaultseppunct}\relax
\EndOfBibitem
\bibitem[Neufeld \latin{et~al.}(2020)Neufeld, Even~Tzur, and
  Cohen]{neufeld2020degree}
Neufeld,~O.; Even~Tzur,~M.; Cohen,~O. Degree of chirality of electromagnetic
  fields and maximally chiral light. \emph{Phys. Rev. A} \textbf{2020},
  \emph{101}, 053831\relax
\mciteBstWouldAddEndPuncttrue
\mciteSetBstMidEndSepPunct{\mcitedefaultmidpunct}
{\mcitedefaultendpunct}{\mcitedefaultseppunct}\relax
\EndOfBibitem
\bibitem[Ayuso \latin{et~al.}(2021)Ayuso, Ordonez, Decleva, Ivanov, and
  Smirnova]{ayuso2021enantio}
Ayuso,~D.; Ordonez,~A.~F.; Decleva,~P.; Ivanov,~M.; Smirnova,~O.
  Enantio-sensitive unidirectional light bending. \emph{Nat. Commun.}
  \textbf{2021}, \emph{12}, 1--9\relax
\mciteBstWouldAddEndPuncttrue
\mciteSetBstMidEndSepPunct{\mcitedefaultmidpunct}
{\mcitedefaultendpunct}{\mcitedefaultseppunct}\relax
\EndOfBibitem
\bibitem[Patterson \latin{et~al.}(2013)Patterson, Schnell, and
  Doyle]{patterson2013enantiomer}
Patterson,~D.; Schnell,~M.; Doyle,~J.~M. Enantiomer-specific detection of
  chiral molecules via microwave spectroscopy. \emph{Nature} \textbf{2013},
  \emph{497}, 475--477\relax
\mciteBstWouldAddEndPuncttrue
\mciteSetBstMidEndSepPunct{\mcitedefaultmidpunct}
{\mcitedefaultendpunct}{\mcitedefaultseppunct}\relax
\EndOfBibitem
\bibitem[Patterson and Doyle(2013)Patterson, and Doyle]{patterson2013sensitive}
Patterson,~D.; Doyle,~J.~M. Sensitive chiral analysis via microwave three-wave
  mixing. \emph{Phys. Rev. Lett.} \textbf{2013}, \emph{111}, 023008\relax
\mciteBstWouldAddEndPuncttrue
\mciteSetBstMidEndSepPunct{\mcitedefaultmidpunct}
{\mcitedefaultendpunct}{\mcitedefaultseppunct}\relax
\EndOfBibitem
\bibitem[Shubert \latin{et~al.}(2014)Shubert, Schmitz, Patterson, Doyle, and
  Schnell]{shubert2014identifying}
Shubert,~V.~A.; Schmitz,~D.; Patterson,~D.; Doyle,~J.~M.; Schnell,~M.
  Identifying enantiomers in mixtures of chiral molecules with broadband
  microwave spectroscopy. \emph{Angew. Chem., Int. Ed.} \textbf{2014},
  \emph{53}, 1152--1155\relax
\mciteBstWouldAddEndPuncttrue
\mciteSetBstMidEndSepPunct{\mcitedefaultmidpunct}
{\mcitedefaultendpunct}{\mcitedefaultseppunct}\relax
\EndOfBibitem
\bibitem[Lobsiger \latin{et~al.}(2015)Lobsiger, Perez, Evangelisti, Lehmann,
  and Pate]{lobsiger2015molecular}
Lobsiger,~S.; Perez,~C.; Evangelisti,~L.; Lehmann,~K.~K.; Pate,~B.~H. Molecular
  structure and chirality detection by Fourier transform microwave
  spectroscopy. \emph{J. Phys. Chem. Lett.} \textbf{2015}, \emph{6},
  196--200\relax
\mciteBstWouldAddEndPuncttrue
\mciteSetBstMidEndSepPunct{\mcitedefaultmidpunct}
{\mcitedefaultendpunct}{\mcitedefaultseppunct}\relax
\EndOfBibitem
\bibitem[Shubert \latin{et~al.}(2016)Shubert, Schmitz, P{\'e}rez, Medcraft,
  Krin, Domingos, Patterson, and Schnell]{shubert2016chiral}
Shubert,~V.~A.; Schmitz,~D.; P{\'e}rez,~C.; Medcraft,~C.; Krin,~A.;
  Domingos,~S.~R.; Patterson,~D.; Schnell,~M. Chiral analysis using broadband
  rotational spectroscopy. \emph{J. Phys. Chem. Lett.} \textbf{2016}, \emph{7},
  341--350\relax
\mciteBstWouldAddEndPuncttrue
\mciteSetBstMidEndSepPunct{\mcitedefaultmidpunct}
{\mcitedefaultendpunct}{\mcitedefaultseppunct}\relax
\EndOfBibitem
\bibitem[Kr{\'a}l and Shapiro(2001)Kr{\'a}l, and Shapiro]{kral2001cyclic}
Kr{\'a}l,~P.; Shapiro,~M. Cyclic population transfer in quantum systems with
  broken symmetry. \emph{Phy. Rev. Lett.} \textbf{2001}, \emph{87},
  183002\relax
\mciteBstWouldAddEndPuncttrue
\mciteSetBstMidEndSepPunct{\mcitedefaultmidpunct}
{\mcitedefaultendpunct}{\mcitedefaultseppunct}\relax
\EndOfBibitem
\bibitem[Kr{\'a}l \latin{et~al.}(2003)Kr{\'a}l, Thanopulos, Shapiro, and
  Cohen]{kral2003two}
Kr{\'a}l,~P.; Thanopulos,~I.; Shapiro,~M.; Cohen,~D. Two-step enantio-selective
  optical switch. \emph{Phys. Rev. Lett.} \textbf{2003}, \emph{90},
  033001\relax
\mciteBstWouldAddEndPuncttrue
\mciteSetBstMidEndSepPunct{\mcitedefaultmidpunct}
{\mcitedefaultendpunct}{\mcitedefaultseppunct}\relax
\EndOfBibitem
\bibitem[Thanopulos \latin{et~al.}(2003)Thanopulos, Kr{\'a}l, and
  Shapiro]{thanopulos2003theory}
Thanopulos,~I.; Kr{\'a}l,~P.; Shapiro,~M. Theory of a two-step enantiomeric
  purification of racemic mixtures by optical means: The D 2 S 2 molecule.
  \emph{J. Chem. Phys.} \textbf{2003}, \emph{119}, 5105--5116\relax
\mciteBstWouldAddEndPuncttrue
\mciteSetBstMidEndSepPunct{\mcitedefaultmidpunct}
{\mcitedefaultendpunct}{\mcitedefaultseppunct}\relax
\EndOfBibitem
\bibitem[Li and Bruder(2008)Li, and Bruder]{li2008dynamic}
Li,~Y.; Bruder,~C. Dynamic method to distinguish between left-and right-handed
  chiral molecules. \emph{Phys. Rev. A} \textbf{2008}, \emph{77}, 015403\relax
\mciteBstWouldAddEndPuncttrue
\mciteSetBstMidEndSepPunct{\mcitedefaultmidpunct}
{\mcitedefaultendpunct}{\mcitedefaultseppunct}\relax
\EndOfBibitem
\bibitem[Vitanov and Drewsen(2019)Vitanov, and Drewsen]{vitanov2019highly}
Vitanov,~N.~V.; Drewsen,~M. Highly efficient detection and separation of chiral
  molecules through shortcuts to adiabaticity. \emph{Phys. Rev. Lett.}
  \textbf{2019}, \emph{122}, 173202\relax
\mciteBstWouldAddEndPuncttrue
\mciteSetBstMidEndSepPunct{\mcitedefaultmidpunct}
{\mcitedefaultendpunct}{\mcitedefaultseppunct}\relax
\EndOfBibitem
\bibitem[Kang \latin{et~al.}(2020)Kang, Shi, Song, and Xia]{kang2020effective}
Kang,~Y.-H.; Shi,~Z.-C.; Song,~J.; Xia,~Y. Effective discrimination of chiral
  molecules in a cavity. \emph{Opt. Lett.} \textbf{2020}, \emph{45},
  4952--4955\relax
\mciteBstWouldAddEndPuncttrue
\mciteSetBstMidEndSepPunct{\mcitedefaultmidpunct}
{\mcitedefaultendpunct}{\mcitedefaultseppunct}\relax
\EndOfBibitem
\bibitem[Wu \latin{et~al.}(2020)Wu, Wang, Han, Wang, Su, Xia, Jiang, and
  Song]{wu2020two}
Wu,~J.-L.; Wang,~Y.; Han,~J.-X.; Wang,~C.; Su,~S.-L.; Xia,~Y.; Jiang,~Y.;
  Song,~J. Two-path interference for enantiomer-selective state transfer of
  chiral molecules. \emph{Phys. Rev. Appl.} \textbf{2020}, \emph{13},
  044021\relax
\mciteBstWouldAddEndPuncttrue
\mciteSetBstMidEndSepPunct{\mcitedefaultmidpunct}
{\mcitedefaultendpunct}{\mcitedefaultseppunct}\relax
\EndOfBibitem
\bibitem[Tutunnikov \latin{et~al.}(2021)Tutunnikov, Xu, Field, Nelson, Prior,
  and Averbukh]{tutunnikov2021enantioselective}
Tutunnikov,~I.; Xu,~L.; Field,~R.~W.; Nelson,~K.~A.; Prior,~Y.; Averbukh,~I.~S.
  Enantioselective orientation of chiral molecules induced by terahertz pulses
  with twisted polarization. \emph{Phys. Rev. Res.} \textbf{2021}, \emph{3},
  013249\relax
\mciteBstWouldAddEndPuncttrue
\mciteSetBstMidEndSepPunct{\mcitedefaultmidpunct}
{\mcitedefaultendpunct}{\mcitedefaultseppunct}\relax
\EndOfBibitem
\bibitem[Gerlach and Stern(1922)Gerlach, and Stern]{gerlach1922experimentelle}
Gerlach,~W.; Stern,~O. Der experimentelle nachweis der richtungsquantelung im
  magnetfeld. \emph{Z. Phys.} \textbf{1922}, \emph{9}, 349--352\relax
\mciteBstWouldAddEndPuncttrue
\mciteSetBstMidEndSepPunct{\mcitedefaultmidpunct}
{\mcitedefaultendpunct}{\mcitedefaultseppunct}\relax
\EndOfBibitem
\bibitem[Hirsch(1999)]{hirsch1999spin}
Hirsch,~J.~E. Spin hall effect. \emph{Phys. Rev. Lett.} \textbf{1999},
  \emph{83}, 1834\relax
\mciteBstWouldAddEndPuncttrue
\mciteSetBstMidEndSepPunct{\mcitedefaultmidpunct}
{\mcitedefaultendpunct}{\mcitedefaultseppunct}\relax
\EndOfBibitem
\bibitem[Li \latin{et~al.}(2007)Li, Bruder, and Sun]{li2007generalized}
Li,~Y.; Bruder,~C.; Sun,~C.~P. Generalized Stern-Gerlach effect for chiral
  molecules. \emph{Phys. Rev. Lett.} \textbf{2007}, \emph{99}, 130403\relax
\mciteBstWouldAddEndPuncttrue
\mciteSetBstMidEndSepPunct{\mcitedefaultmidpunct}
{\mcitedefaultendpunct}{\mcitedefaultseppunct}\relax
\EndOfBibitem
\bibitem[Li and Shapiro(2010)Li, and Shapiro]{li2010theory}
Li,~X.; Shapiro,~M. Theory of the optical spatial separation of racemic
  mixtures of chiral molecules. \emph{J. Chem. Phys.} \textbf{2010},
  \emph{132}, 194315\relax
\mciteBstWouldAddEndPuncttrue
\mciteSetBstMidEndSepPunct{\mcitedefaultmidpunct}
{\mcitedefaultendpunct}{\mcitedefaultseppunct}\relax
\EndOfBibitem
\bibitem[Chen \latin{et~al.}(2020)Chen, Ye, Zhang, and Li]{chen2020enantio}
Chen,~Y.-Y.; Ye,~C.; Zhang,~Q.; Li,~Y. Enantio-discrimination via light
  deflection effect. \emph{J. Chem. Phys.} \textbf{2020}, \emph{152},
  204305\relax
\mciteBstWouldAddEndPuncttrue
\mciteSetBstMidEndSepPunct{\mcitedefaultmidpunct}
{\mcitedefaultendpunct}{\mcitedefaultseppunct}\relax
\EndOfBibitem
\bibitem[Kane and Mele(2005)Kane, and Mele]{kane2005quantum}
Kane,~C.~L.; Mele,~E.~J. Quantum spin Hall effect in graphene. \emph{Phys. Rev.
  Lett.} \textbf{2005}, \emph{95}, 226801\relax
\mciteBstWouldAddEndPuncttrue
\mciteSetBstMidEndSepPunct{\mcitedefaultmidpunct}
{\mcitedefaultendpunct}{\mcitedefaultseppunct}\relax
\EndOfBibitem
\bibitem[Thouless \latin{et~al.}(1982)Thouless, Kohmoto, Nightingale, and
  {den}~Nijs]{thouless1982quantized}
Thouless,~D.~J.; Kohmoto,~M.; Nightingale,~M.~P.; {den}~Nijs,~M. Quantized Hall
  conductance in a two-dimensional periodic potential. \emph{Phy. Rev. Lett.}
  \textbf{1982}, \emph{49}, 405\relax
\mciteBstWouldAddEndPuncttrue
\mciteSetBstMidEndSepPunct{\mcitedefaultmidpunct}
{\mcitedefaultendpunct}{\mcitedefaultseppunct}\relax
\EndOfBibitem
\bibitem[Hasan and Kane(2010)Hasan, and Kane]{hasan2010colloquium}
Hasan,~M.~Z.; Kane,~C.~L. Colloquium: topological insulators. \emph{Rev. Mod.
  Phys.} \textbf{2010}, \emph{82}, 3045\relax
\mciteBstWouldAddEndPuncttrue
\mciteSetBstMidEndSepPunct{\mcitedefaultmidpunct}
{\mcitedefaultendpunct}{\mcitedefaultseppunct}\relax
\EndOfBibitem
\bibitem[Kolodrubetz \latin{et~al.}(2018)Kolodrubetz, Nathan, Gazit, Morimoto,
  and Moore]{kolodrubetz2018topological}
Kolodrubetz,~M.~H.; Nathan,~F.; Gazit,~S.; Morimoto,~T.; Moore,~J.~E.
  Topological floquet-thouless energy pump. \emph{Phys. Rev. Lett.}
  \textbf{2018}, \emph{120}, 150601\relax
\mciteBstWouldAddEndPuncttrue
\mciteSetBstMidEndSepPunct{\mcitedefaultmidpunct}
{\mcitedefaultendpunct}{\mcitedefaultseppunct}\relax
\EndOfBibitem
\bibitem[Mondragon-Shem \latin{et~al.}(2018)Mondragon-Shem, Martin,
  Alexandradinata, and Cheng]{mondragon2018quantized}
Mondragon-Shem,~I.; Martin,~I.; Alexandradinata,~A.; Cheng,~M. Quantized
  frequency-domain polarization of driven phases of matter. \emph{arXiv
  preprint arXiv:1811.10632} \textbf{2018}, \relax
\mciteBstWouldAddEndPunctfalse
\mciteSetBstMidEndSepPunct{\mcitedefaultmidpunct}
{}{\mcitedefaultseppunct}\relax
\EndOfBibitem
\bibitem[Nathan \latin{et~al.}(2019)Nathan, Martin, and
  Refael]{nathan2019topological}
Nathan,~F.; Martin,~I.; Refael,~G. Topological frequency conversion in a driven
  dissipative quantum cavity. \emph{Phys. Rev. B} \textbf{2019}, \emph{99},
  094311\relax
\mciteBstWouldAddEndPuncttrue
\mciteSetBstMidEndSepPunct{\mcitedefaultmidpunct}
{\mcitedefaultendpunct}{\mcitedefaultseppunct}\relax
\EndOfBibitem
\bibitem[Oka and Kitamura(2019)Oka, and Kitamura]{oka2019floquet}
Oka,~T.; Kitamura,~S. Floquet engineering of quantum materials. \emph{Annu.
  Rev. Condens. Matter Phys.} \textbf{2019}, \emph{10}, 387--408\relax
\mciteBstWouldAddEndPuncttrue
\mciteSetBstMidEndSepPunct{\mcitedefaultmidpunct}
{\mcitedefaultendpunct}{\mcitedefaultseppunct}\relax
\EndOfBibitem
\bibitem[Leboeuf \latin{et~al.}(1990)Leboeuf, Kurchan, Feingold, and
  Arovas]{leboeuf1990phase}
Leboeuf,~P.; Kurchan,~J.; Feingold,~M.; Arovas,~D.~P. Phase-space localization:
  topological aspects of quantum chaos. \emph{Phys. Rev. Lett.} \textbf{1990},
  \emph{65}, 3076\relax
\mciteBstWouldAddEndPuncttrue
\mciteSetBstMidEndSepPunct{\mcitedefaultmidpunct}
{\mcitedefaultendpunct}{\mcitedefaultseppunct}\relax
\EndOfBibitem
\bibitem[Leboeuf \latin{et~al.}(1992)Leboeuf, Kurchan, Feingold, and
  Arovas]{leboeuf1992topological}
Leboeuf,~P.; Kurchan,~J.; Feingold,~M.; Arovas,~D.~P. Topological aspects of
  quantum chaos. \emph{Chaos} \textbf{1992}, \emph{2}, 125--130\relax
\mciteBstWouldAddEndPuncttrue
\mciteSetBstMidEndSepPunct{\mcitedefaultmidpunct}
{\mcitedefaultendpunct}{\mcitedefaultseppunct}\relax
\EndOfBibitem
\bibitem[Martin \latin{et~al.}(2017)Martin, Refael, and
  Halperin]{martin2017topological}
Martin,~I.; Refael,~G.; Halperin,~B. Topological frequency conversion in
  strongly driven quantum systems. \emph{Phys. Rev. X} \textbf{2017}, \emph{7},
  041008\relax
\mciteBstWouldAddEndPuncttrue
\mciteSetBstMidEndSepPunct{\mcitedefaultmidpunct}
{\mcitedefaultendpunct}{\mcitedefaultseppunct}\relax
\EndOfBibitem
\bibitem[Ordonez and Smirnova(2019)Ordonez, and
  Smirnova]{ordonez2019propensity}
Ordonez,~A.~F.; Smirnova,~O. Propensity rules in photoelectron circular
  dichroism in chiral molecules. II. General picture. \emph{Phys. Rev. A}
  \textbf{2019}, \emph{99}, 043417\relax
\mciteBstWouldAddEndPuncttrue
\mciteSetBstMidEndSepPunct{\mcitedefaultmidpunct}
{\mcitedefaultendpunct}{\mcitedefaultseppunct}\relax
\EndOfBibitem
\bibitem[Lehmann(2018)]{lehmann2018influence}
Lehmann,~K.~K. Influence of spatial degeneracy on rotational spectroscopy:
  Three-wave mixing and enantiomeric state separation of chiral molecules.
  \emph{J. Chem. Phys.} \textbf{2018}, \emph{149}, 094201\relax
\mciteBstWouldAddEndPuncttrue
\mciteSetBstMidEndSepPunct{\mcitedefaultmidpunct}
{\mcitedefaultendpunct}{\mcitedefaultseppunct}\relax
\EndOfBibitem
\bibitem[Leibscher \latin{et~al.}(2019)Leibscher, Giesen, and
  Koch]{leibscher2019principles}
Leibscher,~M.; Giesen,~T.~F.; Koch,~C.~P. Principles of enantio-selective
  excitation in three-wave mixing spectroscopy of chiral molecules. \emph{J.
  Chem. Phys.} \textbf{2019}, \emph{151}, 014302\relax
\mciteBstWouldAddEndPuncttrue
\mciteSetBstMidEndSepPunct{\mcitedefaultmidpunct}
{\mcitedefaultendpunct}{\mcitedefaultseppunct}\relax
\EndOfBibitem
\bibitem[Simons(2003)]{simons2003introduction}
Simons,~J. \emph{An introduction to theoretical chemistry}; Cambridge
  University Press: Cambridge, 2003\relax
\mciteBstWouldAddEndPuncttrue
\mciteSetBstMidEndSepPunct{\mcitedefaultmidpunct}
{\mcitedefaultendpunct}{\mcitedefaultseppunct}\relax
\EndOfBibitem
\bibitem[Boyers \latin{et~al.}(2020)Boyers, Crowley, Chandran, and
  Sushkov]{boyers2020exploring}
Boyers,~E.; Crowley,~P. J.~D.; Chandran,~A.; Sushkov,~A.~O. Exploring 2D
  Synthetic Quantum Hall Physics with a Quasiperiodically Driven Qubit.
  \emph{Phys. Rev. Lett.} \textbf{2020}, \emph{125}, 160505\relax
\mciteBstWouldAddEndPuncttrue
\mciteSetBstMidEndSepPunct{\mcitedefaultmidpunct}
{\mcitedefaultendpunct}{\mcitedefaultseppunct}\relax
\EndOfBibitem
\bibitem[Samoilenko(2012)]{samoilenko2012elements}
Samoilenko,~A.~M. \emph{Elements of the mathematical theory of multi-frequency
  oscillations}; Springer Science \& Business Media: Berlin, 2012;
  Vol.~71\relax
\mciteBstWouldAddEndPuncttrue
\mciteSetBstMidEndSepPunct{\mcitedefaultmidpunct}
{\mcitedefaultendpunct}{\mcitedefaultseppunct}\relax
\EndOfBibitem
\bibitem[Bernevig and Zhang(2006)Bernevig, and Zhang]{bernevig2006quantum}
Bernevig,~B.~A.; Zhang,~S.-C. Quantum spin Hall effect. \emph{Phys. Rev. Lett.}
  \textbf{2006}, \emph{96}, 106802\relax
\mciteBstWouldAddEndPuncttrue
\mciteSetBstMidEndSepPunct{\mcitedefaultmidpunct}
{\mcitedefaultendpunct}{\mcitedefaultseppunct}\relax
\EndOfBibitem
\bibitem[Schellman(1975)]{schellman1975circular}
Schellman,~J.~A. Circular dichroism and optical rotation. \emph{Chem. Rev.}
  \textbf{1975}, \emph{75}, 323--331\relax
\mciteBstWouldAddEndPuncttrue
\mciteSetBstMidEndSepPunct{\mcitedefaultmidpunct}
{\mcitedefaultendpunct}{\mcitedefaultseppunct}\relax
\EndOfBibitem
\bibitem[Park and Field(2016)Park, and Field]{park2016perspective}
Park,~G.~B.; Field,~R.~W. Perspective: The first ten years of broadband chirped
  pulse Fourier transform microwave spectroscopy. \emph{J. Chem. Phys.}
  \textbf{2016}, \emph{144}, 200901\relax
\mciteBstWouldAddEndPuncttrue
\mciteSetBstMidEndSepPunct{\mcitedefaultmidpunct}
{\mcitedefaultendpunct}{\mcitedefaultseppunct}\relax
\EndOfBibitem
\bibitem[Belkin \latin{et~al.}(2000)Belkin, Kulakov, Ernst, Yan, and
  Shen]{belkin2000sum}
Belkin,~M.; Kulakov,~T.; Ernst,~K.-H.; Yan,~L.; Shen,~Y. Sum-frequency
  vibrational spectroscopy on chiral liquids: a novel technique to probe
  molecular chirality. \emph{Phys. Rev. Lett.} \textbf{2000}, \emph{85},
  4474\relax
\mciteBstWouldAddEndPuncttrue
\mciteSetBstMidEndSepPunct{\mcitedefaultmidpunct}
{\mcitedefaultendpunct}{\mcitedefaultseppunct}\relax
\EndOfBibitem
\bibitem[Fischer \latin{et~al.}(2003)Fischer, Buckingham, Beckwitt, Wiersma,
  and Wise]{fischer2003new}
Fischer,~P.; Buckingham,~A.; Beckwitt,~K.; Wiersma,~D.; Wise,~F. New
  electro-optic effect: sum-frequency generation from optically active liquids
  in the presence of a dc electric field. \emph{Phys. Rev. Lett.}
  \textbf{2003}, \emph{91}, 173901\relax
\mciteBstWouldAddEndPuncttrue
\mciteSetBstMidEndSepPunct{\mcitedefaultmidpunct}
{\mcitedefaultendpunct}{\mcitedefaultseppunct}\relax
\EndOfBibitem
\bibitem[Faure and Zhilinskii(2000)Faure, and Zhilinskii]{faure2000topological}
Faure,~F.; Zhilinskii,~B. Topological Chern indices in molecular spectra.
  \emph{Phys. Rev. Lett.} \textbf{2000}, \emph{85}, 960\relax
\mciteBstWouldAddEndPuncttrue
\mciteSetBstMidEndSepPunct{\mcitedefaultmidpunct}
{\mcitedefaultendpunct}{\mcitedefaultseppunct}\relax
\EndOfBibitem
\bibitem[Schwennicke and Yuen-Zhou(2020)Schwennicke, and
  Yuen-Zhou]{schwennicke2020optical}
Schwennicke,~K.; Yuen-Zhou,~J. Optical activity from the exciton Aharonov--Bohm
  effect: A Floquet engineering approach. \emph{J. Phys. Chem. C}
  \textbf{2020}, \emph{124}, 4206--4214\relax
\mciteBstWouldAddEndPuncttrue
\mciteSetBstMidEndSepPunct{\mcitedefaultmidpunct}
{\mcitedefaultendpunct}{\mcitedefaultseppunct}\relax
\EndOfBibitem
\bibitem[Ordonez \latin{et~al.}(2021)Ordonez, Ayuso, Decleva, and
  Smirnova]{ordonez2021geometric}
Ordonez,~A.~F.; Ayuso,~D.; Decleva,~P.; Smirnova,~O. Geometric magnetism and
  new enantio-sensitive observables in photoionization of chiral molecules.
  \emph{arXiv preprint arXiv:2106.14264} \textbf{2021}, \relax
\mciteBstWouldAddEndPunctfalse
\mciteSetBstMidEndSepPunct{\mcitedefaultmidpunct}
{}{\mcitedefaultseppunct}\relax
\EndOfBibitem
\end{mcitethebibliography}

\end{document}